\documentclass[%
groupedaddress,
preprint,
nobibnotes,
 amsmath,
 amssymb,
 aps,
 citeautoscript,
prl
]{revtex4-1}

\usepackage{graphicx}% Include figure files
\usepackage{dcolumn}% Align table columns on decimal point
\usepackage{bm}% bold math
\usepackage{epstopdf}
\usepackage{epsfig}
\usepackage{booktabs}
\usepackage{array}
\newcolumntype{x}[1]{{\centering\hspace{0pt}}p{#1}}%
\graphicspath{ {Plots/} }
\begin{document}

%\preprint{APS/123-QED}

\title{Formation of two-dimensional electron and hole gases at the interface of half-Heusler semiconductors}% Force line breaks with \\

\author{Abhishek Sharan}
\affiliation{Department of Physics and Astronomy, University of Delaware, Newark, DE 19716}
\author{Zhigang Gui, Anderson Janotti}
\affiliation{Department of Materials Science and Engineering, University of Delaware, Newark, DE 19716}

\date{\today}% It is always \today, today,

\begin{abstract}

Heuslers are a prominent family of multi-functional materials that includes semiconductors, 
half metals, topological semimetals, and magnetic superconductors. Owing to their same crystalline structure, yet quite different electronic properties and flexibility in chemical composition,
Heusler-based heterostructures can be designed to show intriguing properties at the interface. 
Using electronic structure calculations, we show that 
two dimensional electron or hole gases (2DEG or 2DHG) form at the interface of half-Heusler (HH) semiconductors without any chemical doping.
We use CoTiSb/NiTiSn as an example, and show that the 2DEG at the TiSb/Ni(001) termination and the 2DHG  
at the Co/TiSn(001) termination are intrinsic to the interface, and hold rather high charge densities of 3$\times$10$^{14}$ carriers/cm$^{2}$.
These excess charge carriers are tightly bound to the interface plane and are fully accommodated in transition-metal $d$ sub-bands.
The formation of 2DEG and 2DHG are not specific to the CoTiSb/NiTiSn system; a list of combinations of HH semiconductors 
that are predicted to form 2DEG or 2DHG is provided based on band alignment, interface termination, and lattice mismatch.

\end{abstract}

\maketitle

Heusler compounds exhibit many exciting properties, including giant magnetoresistance (GMR) \cite{Pierre1997, Pierre1999}, piezoelectricity \cite{Roy2012}, half-metallic ferromagnetism \cite{DeGroot1983,Tanaka1999, Felser2007}, superconductivity \cite{Tafti2013,Klimczuk2012}, and non-trivial topological electronic structure \cite{Lin2015,Logan2016}. This diversity in properties is intimately linked to their flexibility in chemical composition and the number of valence electrons \cite{Graf2011}. Heuslers with 20 and 21 valence electrons are ferromagnetic \cite{Kandpal2006, Offernes2007, Graf2011}, whereas half-metallic ferromagnetism is generally found in materials with 22 valence electrons. The multi-functional nature and tunability of these Heusler compounds open up vast opportunities in device applications. Half-Heusler (HH) semiconductors, having 8 or 18 valence electrons and band gaps ranging from 0 to 4 eV, form an important sub-class of Heusler compounds, consisting of over 250 ternary compounds. They have attracted wide interest due to their potential application in spintronics \cite{Ma2017}, solar cells \cite{Bechstedt2016}, and thermoelectrics \cite{Zeier2016}. Many HH are lattice matched to and can be epitaxially grown on III-V semiconductor substrates \cite{Harrington2017}, adding great flexibility in device design and integration with conventional semiconductors.  Here, we show that interfaces of HH semiconductors exhibit two-dimensional electron or hole gases (2DEG or 2DHG), without any chemical doping.
Excess charge densities of the order of 3$\times$10$^{14}$ carriers/cm$^2$ are predicted, and is inherent to the polar/nonpolar nature of the interface.

\begin{figure}[h]
\includegraphics[width=8 cm]{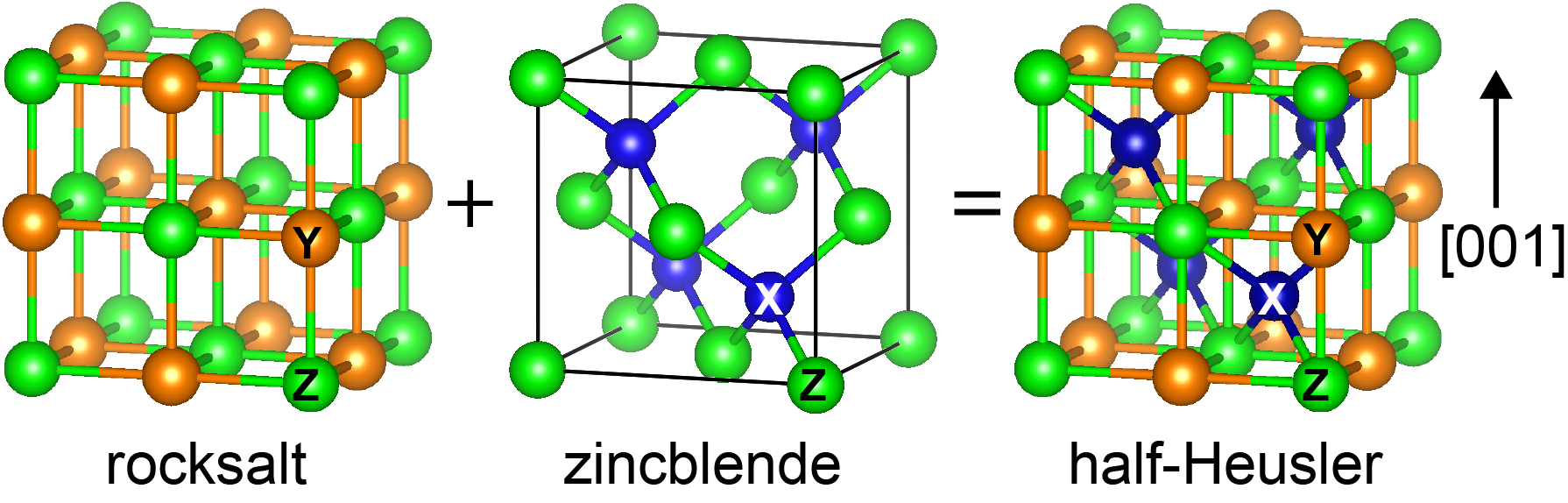}
\caption{Crystal structure of half-Heusler (HH) compounds XYZ.  It can be viewed as interpenetrating rocksalt YZ and zincblende XZ sub-lattices. Along the [001] direction, the structure has alternating planes consisting of X and YZ atoms.}
\label{fig1}
\end{figure}

HH compounds are ternary with composition XYZ, where X and Y are typically transition metals and Z is a main group element. The crystal structure belongs to the space group $F\overline{4}3m$, and consists of three interpenetrating face-centered cubic (fcc) sub-lattices, or, equivalently, interpenetrating zincblende XZ and rocksalt YZ sub-lattices as shown in Fig.~\ref{fig1}. Along the [001] direction, the HH compounds can be viewed as alternating planes of X and YZ atoms. Thus the oxidation state of X, Y and Z, which basically represents the number of electrons lost or gained by an atom of that element in the compound, will determine if the XYZ compound is polar or non-polar along [001]. By combining polar and non-polar HH compounds in heterostructures along [001], with atomically sharp interfaces, we expect the polar mismatch to result in excess mobile charge carriers that will occupy conduction-band states (electrons) or valence-band  states (holes), depending on the interface termination and band alignment between the two HH materials. This excess charge is expected to be bound to the fixed charges at the interface plane, forming two-dimensional electron or hole gas systems.
Through first-principles electronic structure calculations based on the density functional theory \cite{Hohenberg1964,Kohn1965} (see Methods and Supplemental Information), we demonstrate this effect by using CoTiSb and NiTiSn as example, and then provide a list of HH material combinations for which we predict the same effects to occur.

\begin{figure}[h]
\includegraphics[width=8.4 cm]{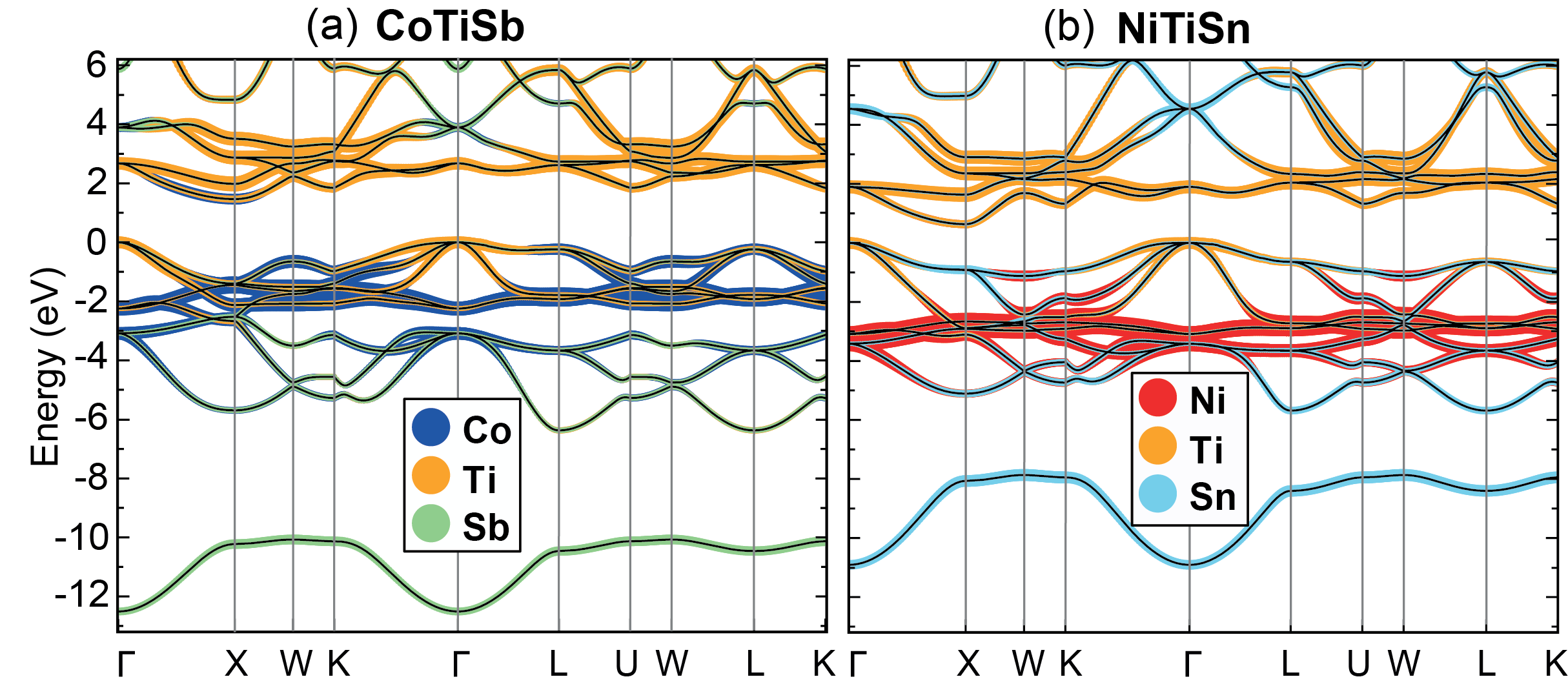}
\caption{Electronic structure of (a) CoTiSb and (b) NiTiSn. Both compounds are indirect-gap semiconductors with valence-band maximum (VBM) at $\Gamma$ and conduction-band minimum (CBM) at the X point. 
The color code represents the contributions from Co, Ni, and Ti $3d$ orbitals, and from $5s$ and $5p$ orbitals in case of Sb and Sn. In both cases, the VBM is set to zero.}
\label{fig2}
\end{figure}

First, we address the polar and non-polar nature of CoTiSb and NiTiSn by inspecting their electronic band structures (shown in Fig.~\ref{fig2}).  The orbital-resolved band structure of CoTiSb [Fig.~\ref{fig2}(a)] shows a band gap separating the occupied valence band, mostly derived from Co 3$d$, from the unoccupied conduction-band, derived from Ti 3$d$. The bands originating from Sb 5$s$ and 5$p$ are fully occupied and located below the Co 3$d$ bands.  From these results we can assign the oxidation states +4 to Ti, -1 to Co, and -3 to Sb, or simply, Ti$^{+4}$, Co$^{-1}$, and Sb$^{-3}$.  Thus, along the [001] direction, CoTiSb can be viewed as composed of alternating negatively charged planes of Co$^{-1}$ and positively charged planes of (TiSb)$^{+1}$.  
 
Similarly, in the case of NiTiSn, the valence band is composed of Ni 3$d$ states, and the conduction band of Ti 3$d$, whereas the Sn 5$s$ and 5$p$ bands are occupied and located below the Ni $3d$ bands, as shown in Fig.~\ref{fig2}(b). Thus, the oxidation states can be assigned as Ti$^{+4}$, Ni$^0$, and Sn$^{-4}$, so that NiTiSn can be viewed as composed of alternating charge neutral planes of Ni$^0$ and (TiSn)$^{0}$.  

\begin{figure}[h]
\includegraphics[width= 7 cm]{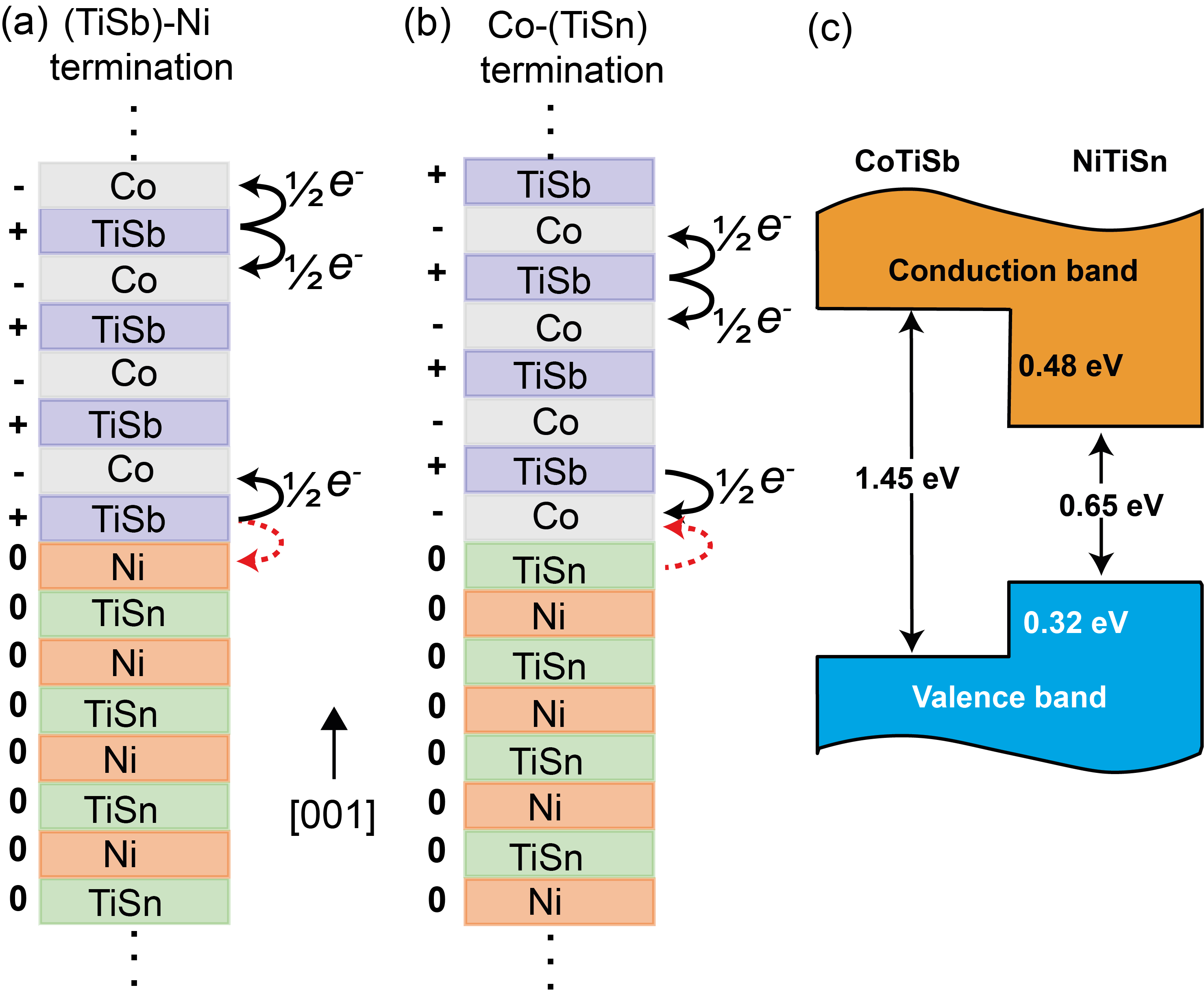}
\caption{Formation of 2DEG and 2DHG at the (001) interface between polar (CoTiSb) and non-polar (NiTiSn) half-Heusler semiconductors.
Both (a) (TiSb)-Ni terminated and (b) Co-(TiSn) terminated interfaces are shown.  Net charges on each layer are indicated. The arrows represent the transfer of electrons between the layers. Red dashed arrow represents excess or missing electrons. Because of polar mismatch at the interface there are excess electrons (2DEG) in the case of (TiSb)-Ni termination, and excess holes (2DHG) in case of Co-(TiSn) interface termination.
(c) Band alignment between CoTiSb and NiTiSn, indicating that excess electrons or holes from the interface will reside mainly on the NiTiSn side.
}
\label{fig3}
\end{figure}

If a heterojunction is made of CoTiSb and NiTiSn along the [001] direction, excess mobile charge will result from the polar mismatch, as illustrated in Fig.~\ref{fig3}. For instance, for an atomically sharp interface of CoTiSb/NiTiSn(001) with (TiSb)-Ni termination [Fig.~\ref{fig3}(a)], there will be an excess of 1/2 electron per unit-cell area, or 3$\times$10$^{14}$ electrons/cm$^2$. 
On the other hand, for the Co-(TiSn) termination, 1/2 electron per unit-cell area will be missing, i.e., there will be excess holes with a density of 3$\times$10$^{14}$ holes/cm$^2$, as illustrated in Fig.~\ref{fig3}(b). 
Where this rather high density of excess electrons or holes will be accommodated depends on the relative position of the valence and conduction bands of the two materials, i.e., their band alignment, and respective density of states near the band edges. In any case, the excess charge will be bound to the plane of fixed charges at the interface, forming a 2DEG or 2DHG.  Note that here we are neglecting any effect of nearby surfaces that may act as source or sink of carriers due to unpassivated dangling bonds.

Knowing the band alignment is crucial to understand the electronic properties of any heterojunction.  The band alignment can be calculated using periodic supercells through the following procedure \cite{Bjaalie2014}: first we determine the valence-band maximum (VBM) and conduction-band minimum (CBM) of each bulk material with respect to the respective average electrostatic potential, and then determine the difference in the average electrostatic potentials using a superlattice geometry where the thickness of each material is enough to converge the average electrostatic potentials in the bulk regions of the superlattice. In our case, we took care to separate the effect of excess charges due to the interface termination from the intrinsic alignment between the two materials (natural band offset).  For this purpose, we chose a superlattice along the non-polar [110] direction, where each plane parallel to the interface contains one formula unit of the same material (details in Methods and Supplemental Information).  The resulting band alignment between CoTiSb and NiTiSn is shown in Fig.~\ref{fig3}(c).
The VBM of NiTiSn is 0.32 eV higher than that of CoTiSb, whereas the CBM is 0.48 eV lower than in CoTiSb, forming a type I, straddling gap alignment.  Therefore, we expect that most of the excess electrons and holes to be located on the NiTiSn side of the interface. However, we anticipate that the relatively small valence-band and conduction-band offsets suggest that spillover of the excess charge to the CoTiSb side may be expected, yet the carriers will still be bound to the interface plane, forming a 2D system.

%%\begin{figure}[h]
%%\includegraphics[width= 3 cm]{Fig4}
%%\caption{Band alignment between CoTiSb and NiTiSn. Both valence-band and conduction-band offsets are shown, indicating that excess electrons or holes from the interface will reside mainly on the NiTiSn side.
%%The calculation for the superlattice was performed using a superlattice along the [110] direction to avoid the buildup charge and potential slope in the bulk regions due to the polar mismatch at the (001) interface.
%%}
%%\label{fig4}
%%\end{figure}

\begin{figure}[h!]
\includegraphics[width=6 cm]{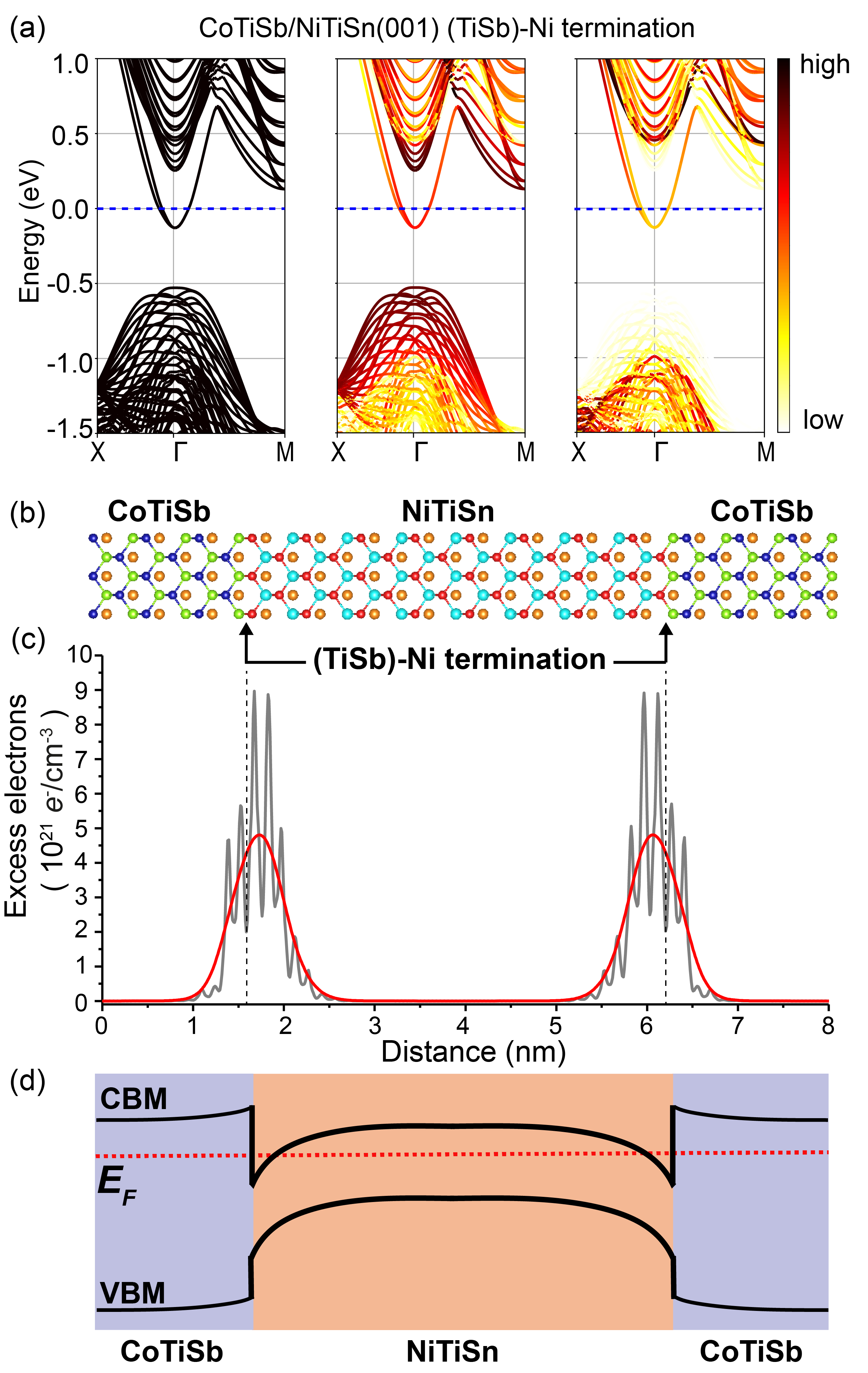}
\caption{Electronic structure and excess electron distribution in the (TiSb)-Ni terminated CoTiSb/NiTiSn(001) superlattice. In (a), the Fermi level ($E_F$, set to zero) crosses electron bands, indicating metallic behavior.
The contributions from the $d$ orbitals of Ni and Ti atoms on the NiTiSn side of the interface are shown in the middle panel, and from the $d$ orbitals of the Co and Ti atoms on the CoTiSb side are shown on the right panel.  (b) (TiSb)-Ni terminated CoTiSb/NiTiSn(001) superlattice, with Co atoms in dark blue, Ni in red, Ti in orange, Sb in green and Sn in light blue. (c) Planar and macroscopically averaged charged density of the occupied electron sub-bands, from CBM to $E_F$. (d) Schematic band diagram for the superlattice, indicating the expected position of $E_F$ with respect to the CBM and VBM across the superlattice.
}
\label{fig4}
\end{figure}

\begin{figure}
\includegraphics[width=6 cm]{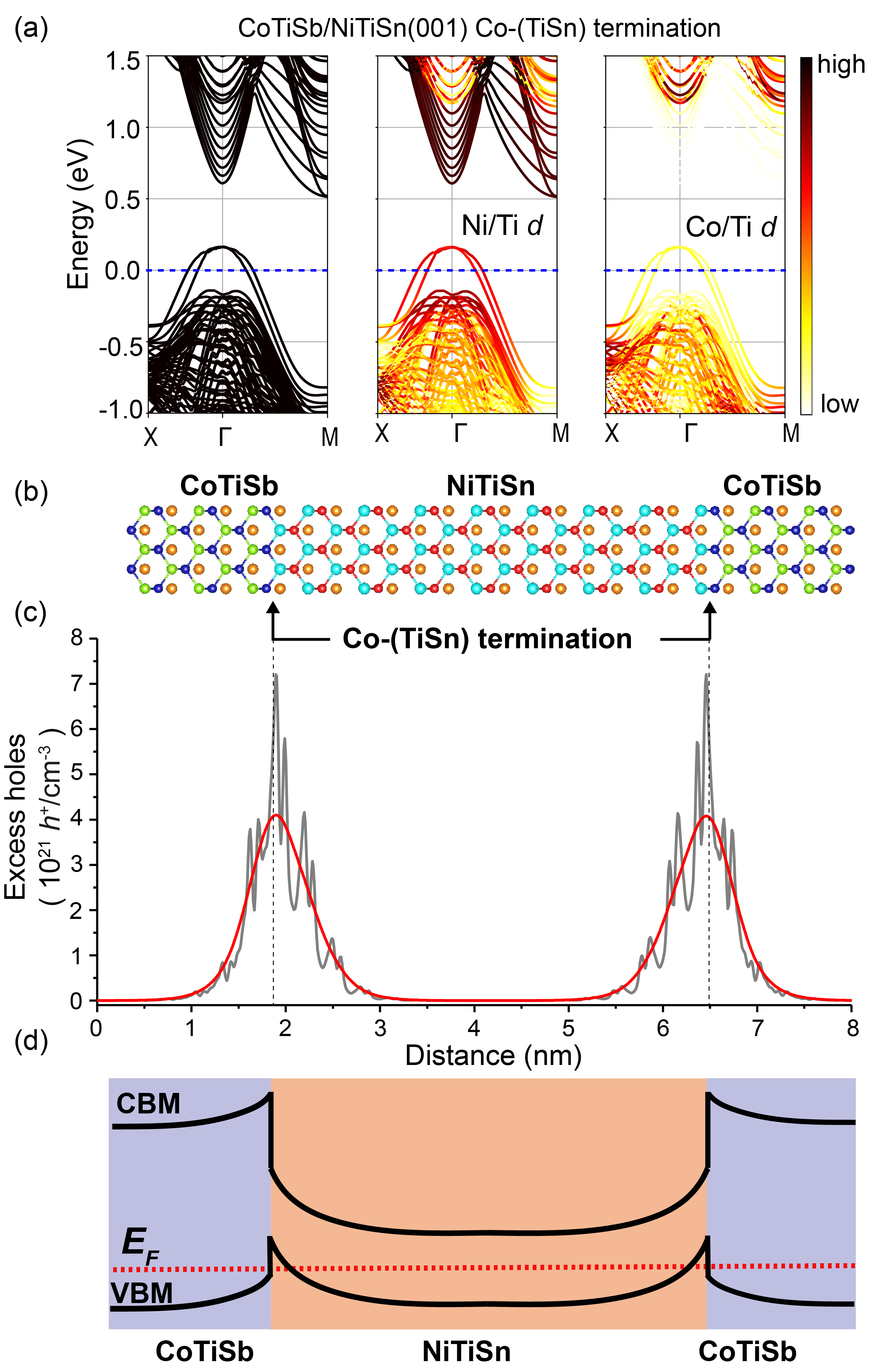}
\caption{Electronic structure and excess charge distribution in the Co-(TiSn) terminated CoTiSb-NiTiSn(001) superlattice. In (a), the Fermi level ($E_F$, set to zero), crosses hole bands, indicating metallic behavior.  The contributions from the $d$ orbitals of Ni and Ti atoms on the NiTiSn side of the interface are shown in the middle panel, and from the $d$ orbitals of the Co and Ti atoms on the CoTiSb side are shown on the right panel. (b) Co-(TiSn) terminated CoTiSb-NiTiSn(001) superlattice,  with Co atoms in dark blue, Ni in red, Ti in orange, Sb in green and Sn in light blue. (c) Planar and macroscopically averaged charged density of the unoccupied hole sub-bands, from VBM down to $E_F$. (d) Schematic band diagram for the superlattice, indicating the expected position of of $E_F$ with respect to the CBM and VBM across the superlattice.
}
\label{fig5}
\end{figure}

To demonstrate the formation of a 2DEG and a 2DHG at the CoTiSb/NiTiSn(001) superlattice interface, we carried out calculations for the electronic structure of superlatices with two equivalent interfaces, i.e., a CoTiSb/NiTiSn(001) superlattice with two (TiSb)-Ni interfaces for the case of excess electrons, shown in Fig.~\ref{fig4}, and a CoTiSb/NiTiSn(001) superlattice with two Co-(TiSn) interfaces for the case of excess holes, shown in Fig.~\ref{fig5}. In either case, the only difference between the two interfaces in each superlattice is that the bonds at one interface are rotated by 90$^{\circ}$ with respect to the bonds at the other interface due to the symmetry of the underlying zincblende sub-lattice. The superlattices consists of 25 layers of CoTiSb (barrier) and 31 layers of NiTiSn (well).  Note that we only used superlattice configurations with two equivalent interfaces for computational convenience of using periodically repeated supercells.  Having two equivalent interfaces is necessary to avoid any charge transfer across the bulk materials and the consequent charge compensation. As a result, the total excess charge in the superlattice will be twice the excess charge due to one interface. 

The electronic structure of the CoTiSb/NiTiSn(001) superlattice with two (TiSb)-Ni interfaces is shown in Fig.~\ref{fig4}(a). The Fermi level is located in the conduction band due to the presence of excess electrons. The orbital-resolved band structures indicate that the occupied conduction-band states are mostly from the Ni and Ti atoms located at or near the interface and on the NiTiSn side.  There is a smaller yet sizable contribution from Co and Ti atoms from CoTiSb.  The charge density distribution corresponding to these occupied conduction bands, averaged in the plane and plotted as a function of the coordinate along the superlattice [001] direction, shown in Figure~\ref{fig4}(b), indicates that the excess electrons are in fact bound to the interface plane, forming a 2DEG.  These excess electrons are accommodated in a region that is 1.5 nm thick at the interface, equivalent to 12 atomic layers.
The amount of charge spillover on the CoTiSb side of the interface depends on a combination of effects, such as conduction-band offset, the thickness of the well (NiTiSn layer), and the density of states of the two compounds.  We also note that the occupied conduction bands are quite dispersive, and we expect electrons with high mobility along the plane of the interface.

The electronic structure of the CoTiSb/NiTiSn(001) superlattice with two Co-(TiSn) interfaces is shown in Fig.~\ref{fig5}(a). The Fermi level is now located in the valence band due the presence of excess holes in the system. The orbital-resolved band structures indicate that the unoccupied valence band states are mostly from the Ni and Ti atoms located at or near the interface, on the NiTiSn side.   The charge density distribution corresponding to these unoccupied valence bands, averaged in the plane and plotted as a function of the coordinate along the superlattice [001] direction, shown in Fig.~\ref{fig5}(b), indicates that the holes are also tightly bound to the interface plane, forming a 2DHG. We note a larger spillover of holes (compared with electrons) to the CoTiSb side due to the relatively small valence-band offset.  The valence bands are much less dispersive than the conduction bands in Fig.~\ref{fig4}(a) suggesting that less the excess holes are less mobile than the excess electrons electrons in the (TiSb)-Ni interface termination.  A schematic representation of the band diagrams for (TiSb)-Ni and Co-(TiSb) interface terminated CoTiSb/NiTiSn (001) superlattices are shown in Fig.~\ref{fig4}(d) and Fig.~\ref{fig5}(d) respectively.

For both Co-(TiSn) and (TiSb)-Ni interface terminations, the integrated excess charge density is 3$\times$10$^{14}$ carriers/cm$^{-2}$ per interface.  This corresponds exactly to 1/2 electron (or hole) per
unit-cell area, and from our assumptions and calculations discussed above, this excess charge is inherent to the interface, i.e., it does not involve any chemical doping or dangling bonds. 
We also note that the excess electrons or holes are strongly bound to the interface by inspecting the charge density profiles in Fig.~\ref{fig4}(b) and ~\ref{fig5}(b).  
\begin{figure}
\includegraphics[width= 5 cm]{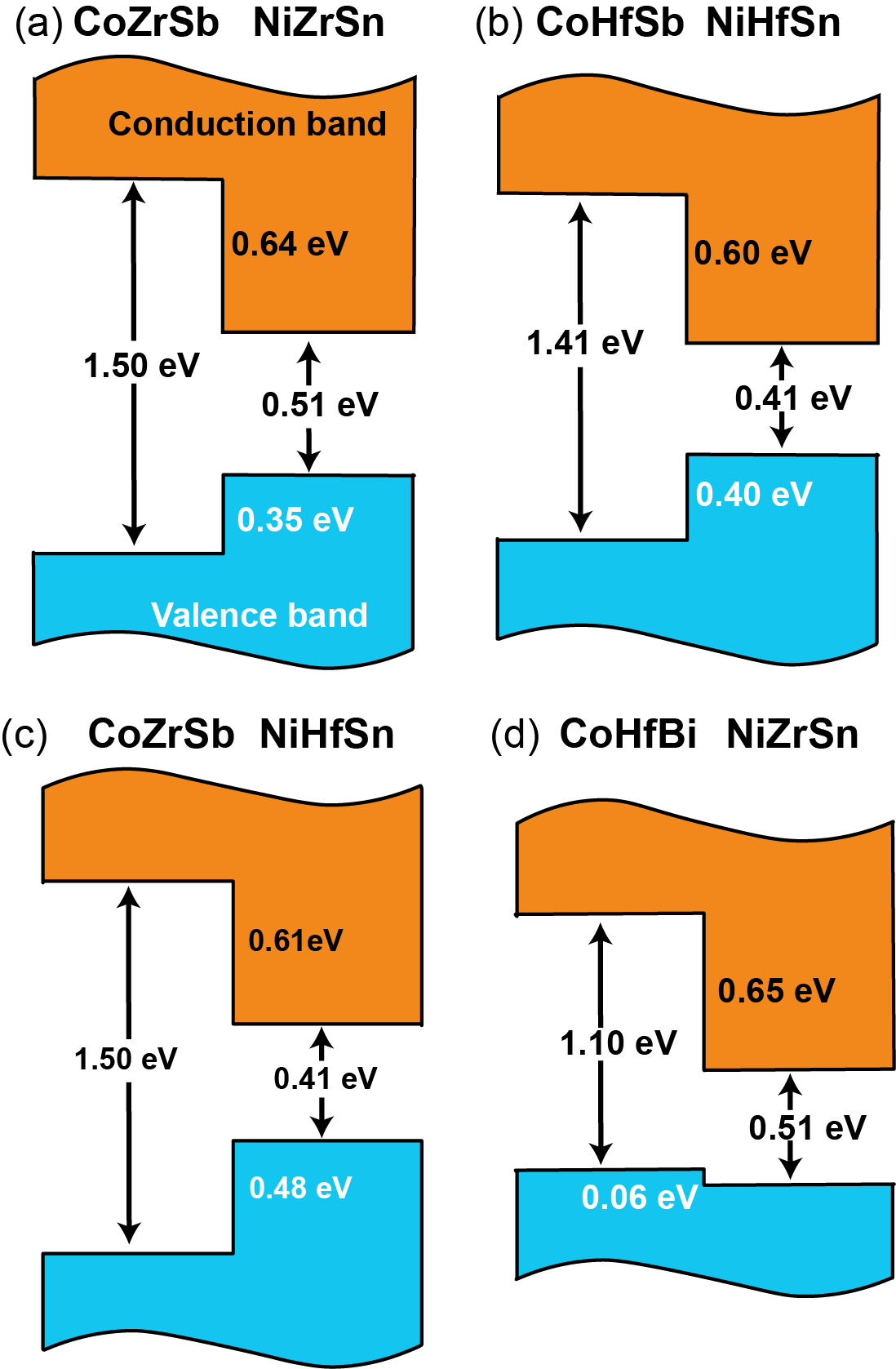}
\caption{Band alignment between (a) CoZrSb and NiZrSn, (b) CoHfSb and NiHfSn, (c) CoZrSb and NiHfSn, and (d) CoHfBi and NiZrSn. }
%Larger conduction band offsets (or valence band offsets) leads to more localization of 2DEG (or 2DHG) at the interface.}
\label{fig6}
\end{figure}
The results for the CoTiSb/NiTiSn interface are quite general and valid for many other combinations of HH compounds (see Fig.~\ref{fig6} and Supplemental Information). For example, for CoZrSb/NiZrSn(001) interface, we have a larger conduction band offset than in CoTiSb/NiTiSn(001), so it is expected that the 2DEG will be more confined to the well in NiZrSn, and perhaps have higher mobility due to the more delocalized nature of the Zr $4d$ bands compared to Ti $3d$ bands.
CoHfSb/NiHfSn(001) and CoZrSb/NiHfSn(001) offer both larger conduction-band and valence-band offsets, with the 2DEG or the 2DHG on the NiHfSn side.  On the other hand, CoHfBi/NiHfSn offers strong confinement for electrons but weak confinement for holes.

Note that the formation of 2DEG due to polar mismatch has been observed at complex-oxide interfaces, such as in SrTiO$_3$/LaAlO$_3$ \cite{Ohtomo2004} and SrTiO$_3$/GdTiO$_3$ \cite{Moetakef2011}. There, the 2DEG is much more spread, likely due to stronger lattice relaxations near the interface and the very high dielectric screening of SrTiO$_3$.  Only recently the existence of a 2DHG has been demonstrated in the case of SrTiO$_3$/LaAlO$_3$ due to the difficulty in controlling the formation of defects that act to compensate holes \cite{Lee2018}. Since, in HH compounds, the valence bands are not too low, and the conduction bands are not too high in an absolute energy scale, i.e., they are comparable to III-V semiconductors \cite{Harrington2017},  we expect that compensation will not be a difficult problem to achieving both 2DEGs and 2DHGs in HH interfaces, as long as sharp interfaces are fabricated.

\section*{Methods}
The calculations are based on the generalized Kohn-Sham theory \cite{Hohenberg1964, Kohn1965} with the Heyd, Scuseria, and Ernzerhof hybrid functional (HSE06) \cite{Heyd2003, Heyd2006} as implemented in VASP code \cite{Kresse1996, Kresse1996a}. The interactions between the valence electrons and the ionic cores are treated using projector-augmented wave potentials \cite{Blochl1994, Kresse1999}. In the HSE06 hybrid functional, the exchange potential is separated into long-range and short-range parts. In the short range part, 25\% of non-local Hartree-Fock exchange is mixed with 75\% of semilocal exchange in the generalized gradient approximation form of Perdew, Burke, and Ernzerhof (PBE) \cite{Perdew1996}. The correlation potential and the long-range part of the exchange are described by PBE. All the calculations were performed using a 350 eV energy cut off for the plane wave expansions. The bulk calculations were performed using a primitive cell of 3 atoms, and a 6$\times$6$\times$6 Monkhorst-Pack special $k$ points for integrations over the Brillouin zone.

The band alignment calculations used superlattices along [110] direction to avoid any excess charge due to the polar mismatch at the interface, since planes along the [110] direction are charge neutral. In these calculations we used superlattice with 11 layers of CoTiSb and 13 layers of NiTiSn, with 2$\times$2$\times$1 Monkhorst-Pack special $k$ points. The methodology followed for calculating band alignments is described in detail in the literature\cite{Bjaalie2014}.

For the superlattices along the [001] direction, to study the formation of 2DEG and 2DHG at the CoTiSb/NiTiSn interface, we used superlattices with 25 layers of CoTiSb and 31 layers of NiTiSn, with 2$\times$2$\times$1 $\Gamma$ centered mesh of k-points. In this case, we have two equivalent interfaces (but rotated, because of symmetry of the zincblende sub-lattice) in the same supercell. The atoms within 8 layers near the interface were fully relaxed, while atoms further away from the interface were fixed to their bulk positions.

For all the HSE06 calculations, we also performed tests using the DFT-GGA functional, to make sure our conclusions are independent of the functional used.  All the conclusions remain unchanged, and the results of such tests are included in the Supplemental information.

%\section*{Acknowledgement}
This worked was supported by the U.S. Department of Energy Basic Energy Science program (DE-SC0014388). It made use of the Extreme Science and Engineering Discovery Environment (XSEDE), which is supported by National Science Foundation grant number ACI-1053575, and the Information Technologies (IT) resources at the University of Delaware, specifically the high-performance computing resources.

%\bibliography{HH}

\end{document}